\documentclass[%
 reprint,
superscriptaddress,
amsmath,amssymb,
prapplied
floatfix,
]{revtex4-2}

\usepackage{graphicx}
\usepackage{dcolumn}
\usepackage{bm}
\usepackage{xcolor}

\usepackage{bbold}
\usepackage{xparse}
\usepackage{hyperref}
\usepackage{siunitx}
\DeclareSIUnit{\amagat}{amg}
\DeclareSIUnit{\sample}{Sa}

\usepackage[nameinlink]{cleveref}
\usepackage{float}
\usepackage{braket,bigints} 

\usepackage{bbm} 

\usepackage{lineno}

\renewcommand\bra[1]{{\langle{#1}|}}
\makeatletter
\renewcommand\ket[1]{%
\@ifnextchar\bra{\k@t{#1}\!}{\k@t{#1}}%
}
\newcommand\k@t[1]{{|{#1}\rangle}}
\makeatother

\usepackage{color}
\definecolor{mygreen}{rgb}{0,0.5,0}
\definecolor{mygrey}{rgb}{0.5,0.5,0.5}
\definecolor{myred}{rgb}{0.75,0,0}
\definecolor{myblue}{rgb}{0,0,0.75}
\definecolor{mymagenta}{cmyk}{0,1,0,0.12}
\definecolor{mycyan}{cmyk}{1,0,0,0.12}
\definecolor{myorange}{rgb}{1.,0.5,0}
\definecolor{myviolet}{rgb}{0.6,0.15,0.6}
\definecolor{mybrown}{cmyk}{0,0.50,1,0.41}

\usepackage{ulem}

\newcommand{\supzero}{^{(0)}}

\newcommand{\subFSR}{_{\mathrm{FSR}}}
\newcommand{\FSR}{\Delta\nu\subFSR}

\newcommand{\subres}{_{\mathrm{res}}}
\newcommand{\omegac}{\omega_0}
\newcommand{\omegares}{\omega\subres}

\newcommand{\tcav}{t_\mathrm{cav}}
\newcommand{\rcav}{r_\mathrm{cav}}
\newcommand{\tcell}{t_\mathrm{cell}}
\newcommand{\OD}{\mathcal{D}}
\newcommand{\twin}{t_\mathrm{win}}
\newcommand{\PolAdded}{a_\mathrm{OP}}
\newcommand{\subpump}{_\mathrm{pump}}
\newcommand{\subprobe}{_\mathrm{probe}}

\newcommand{\supDone}{^{\Done}}
\newcommand{\supDtwo}{^{\Dtwo}}

\newcommand{\PolEffect}{\tilde{\alpha}^{(1)}}

\usepackage{mathtools}
\newcommand{\Dtwo}{\ensuremath{\mathrm{D}_2}{}}
\newcommand{\Done}{\ensuremath{\mathrm{D}_1}{}}
\renewcommand{\Dtwo}{\relax\ifmmode\mathrm{D}_2\else D\textsubscript{2}{}\fi}
\renewcommand{\Done}{\relax\ifmmode\mathrm{D}_1\else D\textsubscript{1}{}\fi}
\newcommand{\Pat}{P_\mathrm{at}}
\renewcommand{\Pat}{S_z}
\newcommand{\Lat}{L_\mathrm{at}}
\newcommand{\Lcav}{L_\mathrm{cav}}

\newcommand{\tat}{t_\mathrm{A}}
\renewcommand{\nat}{n_\mathrm{A}}
\newcommand{\cE}{\mathcal{E}}
\newcommand{\Popt}{P_\mathrm{opt}}
\newcommand{\omegaA}{\omega_\mathrm{at}}

\newcommand{\PLaser}{P_\mathrm{las}}

\newcommand{\fosc}{f_\mathrm{osc}}

\newcommand{\Psat}{P_\mathrm{sat}}
\newcommand{\Ppump}{P_\mathrm{pump}}
\newcommand{\Isat}{I_\mathrm{sat}}
\newcommand{\Ipump}{I_\mathrm{pump}}
\newcommand{\ROP}{R_\mathrm{OP}}
\newcommand{\RRel}{R_\mathrm{rel}}
\newcommand{\OmegaPump}{\Omega\subpump}
\newcommand{\nuPump}{\nu\subpump}
\newcommand{\supRMS}{^{\mathrm{rms}}}
\newcommand{\supPP}{^{\mathrm{P\mbox{-}P}}}
\newcommand{\subsig}{_{\mathrm{sig}}}
\newcommand{\PDone}{\relax\ifmmode\mathrm{PD}_1\else PD\textsubscript{1}{}\fi}
\newcommand{\PDtwo}{\relax\ifmmode\mathrm{PD}_2\else PD\textsubscript{2}{}\fi}

\newcommand{\supRBW}{^{\mathrm{RBW}}}
\newcommand{\RBW}{\Delta\nu\supRBW}

\newcommand{\supREF}{^{\mathrm{ref}}}
\newcommand{\PSA}{P_\mathrm{SA}}
\newcommand{\PSARef}{\PSA\supREF}

\newcommand{\Gammacav}{\Gamma_\mathrm{cav}}

\newcommand{\ICFO}{ICFO - Institut de Ci\`encies Fot\`oniques, The Barcelona Institute of Science and Technology, 08860 Castelldefels (Barcelona), Spain}
\newcommand{\ICREA}{ICREA - Instituci\'{o} Catalana de Recerca i Estudis Avan{\c{c}}ats, 08010 Barcelona, Spain}
\newcommand{\BARI}{Dipartimento Interateneo di Fisica, Universit\'{a}
degli Studi di Bari Aldo Moro, 70126 Bari, Italy}
\newcommand{\China}{School of Mechanical Engineering, Xi’an Jiaotong University, 710049 Xi’an, China}
\newcommand{\Italy}{CNR-INO, Istituto Nazionale di Ottica, 50019 Sesto Fiorentino, Italy}
\begin{document}

\preprint{APS/123-QED}
\author{Mar\'{i}a Hern\'{a}ndez Ruiz}

\affiliation{\ICFO}
\author{Yintao Ma}
\affiliation{\ICFO}
\affiliation{\China}
\author{Hana Medhat}
\affiliation{\ICFO}
\author{Chiara Mazzinghi}
\affiliation{\ICFO}
\affiliation{\Italy}

\author{Vito Giovanni Lucivero} 
\affiliation{\ICFO}
\affiliation{\BARI}
\author{Morgan W. Mitchell}
\email{morgan.mitchell@icfo.eu}
\affiliation{\ICFO}
\affiliation{\ICREA}

\newcommand{\theTitle}{Pound-Drever-Hall detection of  spin polarization in a microfabricated vapor cell}%

\renewcommand{\theTitle}{A microfabricated atomic vapor cell in a resonant optical cavity: Optical pumping and Pound-Drever-Hall detection of spin polarization. }%

\renewcommand{\theTitle}{Cavity-based dispersive detection of spin polarization in a microfabricated vapor cell} 

\renewcommand{\theTitle}{Cavity-enhanced detection of spin polarization in a microfabricated atomic vapor cell}

\title{\theTitle}%

\date{\today}

\begin{abstract}
We demonstrate continuous Pound-Drever-Hall (PDH) nondestructive monitoring of the electron spin polarization of an atomic vapor in a microfabricated vapor cell within an optical resonator. The two-chamber silicon and glass cell contains \textsuperscript{87}Rb and \SI{1.3}{\amagat} of N\textsubscript{2} buffer gas, and is placed within a planar optical resonator formed by two mirrors with dichroic dielectric coatings to resonantly enhance the coupling to phase-modulated probe light near the \Dtwo{} line at \SI{780}{\nano\meter}. We describe the theory of signal generation in this system, including the spin-dependent complex refractive index, cavity optical transfer functions, and PDH signal response to spin polarization. We observe cavity transmission and PDH signals across $\approx \SI{200}{\giga\hertz}$ of detuning around the atomic resonance line. By resonant optical pumping on the \SI{795}{\nano\meter} \Done{} line, we observe spin-dependent cavity line shifts, in good agreement with theory. We use the saturation of the line shift vs.~optical pumping power to calibrate the number density and efficiency of the optical pumping. In the unresolved sideband regime, we observe quantum-noise-limited PDH readout of the spin polarization density, with a flat noise floor of \SI{9e9}{spins\per\centi\meter\cubed\per\sqrt\hertz} for frequencies above \SI{700}{\hertz}. We note possible extensions of the technique. 

\end{abstract}

\maketitle


\section{Introduction}
\label{sec:Introduction} 

Sensing and metrology instruments based on optical pumping and atomic vapor spin physics \cite{Cohen-TannoudjiPiO1966, HapperRMP1972}, including atomic clocks \cite{BellIRETMTT1959} and sensors such as magnetometers \cite{Dupont-RocPLA1969}, gyroscopes \cite{KornackPRL2005, WalkerBook2016}, and systems to test fundamental physics \cite{SafronovaRMP2018}, offer high stability and sensitivity referenced to universally-reproducible atomic spectral features. Many such instruments employ alkali metal atoms such as rubidium or cesium, in combination with methods such as buffer gas \cite{DickePR1953} or anti-relaxation coatings \cite{BouchiatPR1966} to reduce the depolarizing effects of wall collisions. Miniaturized vapor cells made with silicon micro-electromechanical systems (MEMS) techniques\cite{KarlenOE2017, Kitching2018, Dyer2023} or with femtosecond laser writing \cite{Lucivero2022LWVC} enable the miniaturization of such devices \cite{Griffith2010, Cipolletti2021, JimenezBook2017, Hummon2018, Hunter2023}, while preserving the long coherence times and high atomic number densities that underlie their excellent performance.

Optical readout of vapor-based instruments complements optical pumping for spin control, and the quality of this readout has important effects on both sensitivity \cite{TroullinouPRL2021, TroullinouPRL2023} and systematic effects \cite{LuciveroPRAppl2021}. With macrosopic cells, multi-pass geometries \cite{Cooper2016, Limes2020, ShengPRL2013, Li2011} have been used to enhance the effective length of the atomic medium, with a corresponding improvement in the ratio of information-carrying forward scattering to spin-randomizing side-scattering \cite{BaragiolaPRA2014}. Reduced dimensions limit the feasibility of this approach with MEMS cells. 

An alternative to multi-pass methods, suitable for atomic volumes down to the wavelength scale \cite{HungerNJP2010}, is the use of resonant structures, generically called ``cavities.''  Like multi-pass methods, these maintain the probe photons in contact with the atomic medium longer than is possible in single-pass configurations. They moreover take advantage of the Purcell effect \cite{PurcellPR1946} of cavity quantum electrodynamics, which reduces side-scattering and thus depolarization of the atomic medium.
Resonant cavities have been used in precision measurement of atomic properties, for example for non-destructive readout of atomic population in atomic clocks \cite{LodewyckPRA2009, AbgrallCRP2015}, and for quantum non-demolition measurement of cold-atom spin states using heterodyne \cite{BohnetNPhot2014} or homodyne \cite{Hosten2016, CoxPRL2016} detection of the phase of the reflected light. Optical resonators have been used to enhanced Faraday rotation in unpolarized \cite{Mazzinghi2021} and polarized \cite{  Crepaz2015, VasilakisNPhys2015, Chatzidrosos2017} spin ensembles. Approaches include both resonant, dichroism-based \cite{Crepaz2015, Sycz_Gawlik_Zachorowski_2010} and non-resonant, birefringence-based \cite{VasilakisNPhys2015, Mazzinghi2021} probing. Calculations of cavity-enhanced  Faraday rotation in cavity transmission 
\cite{Ling1994, Mazzinghi2021} predict enhancement of rotation angle \cite{Ling1994} and other figures of merit \cite{Mazzinghi2021} for dichroism- \cite{Sycz_Gawlik_Zachorowski_2010} and birefringence-based \cite{Ling1994, Mazzinghi2021} rotation.

\begin{figure*}[t]
	\centering
	\includegraphics[width=0.90\textwidth]{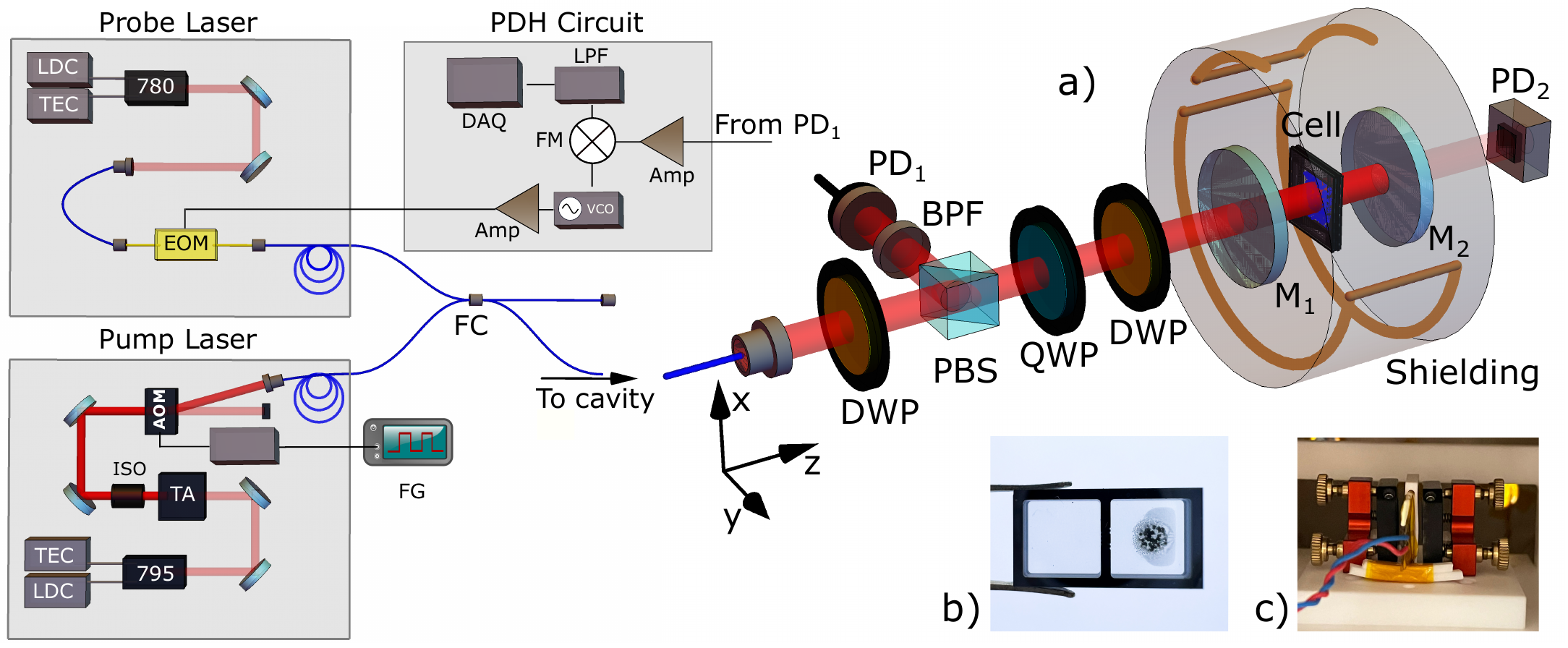}
	\caption{Experimental setup for cavity-based detection of atomic polarization. Pump and probe beams are coupled into a fiber coupler (FC) and reach the atomic vapor within the cavity in the same collinear mode. The probe light is modulated in phase, the pump light is amplified and modulated in amplitude. The reflected probe is collected by a 125-MHz bandwith  photodetector \PDone, whose output is fed into a Pound-Drever-Hall (PDH) circuit. A secondary detector \PDtwo{} collects the transmitted light from the cavity. LDC - laser diode current controller, TEC - laser diode temperature controller, EOM - electro-optic modulator, TA - tapered amplifier, ISO - optical isolator, AOM - acousto-optic modulator, RFD - radio frequency driver, FG - function generator, VCO - voltage-controlled oscillator, FM - frequency mixer, LPF - low-pass filter, DAQ - digital oscilloscope. (a) Sketch of the cavity with detection in reflection. DWP - dual wavelength wave-plate, PBS - polarizing beam splitter, QWP - quarter wave-plate, BPF - laser line bandpass filter, PD- photo-detector, M$_{1/2}$ - first/second planar cavity mirror. (b) MEMS cell. Picture of the two-chamber microfabricated vapor cell after activation by UV decomposition of rubidium azide $^{87}$RbN$_3$. The physics (reservoir) chamber is in the left (right) portion of the cell. (c) Cavity and cell. Picture of the probe resonant cavity consisting of two planar mirrors surrounding the MEMS cell, which is heated and thermally insulated. 
	 }
\label{fig:PDHExpSetup}
\end{figure*}

\begin{figure*}[t]
    \centering
     \includegraphics[scale=0.38]{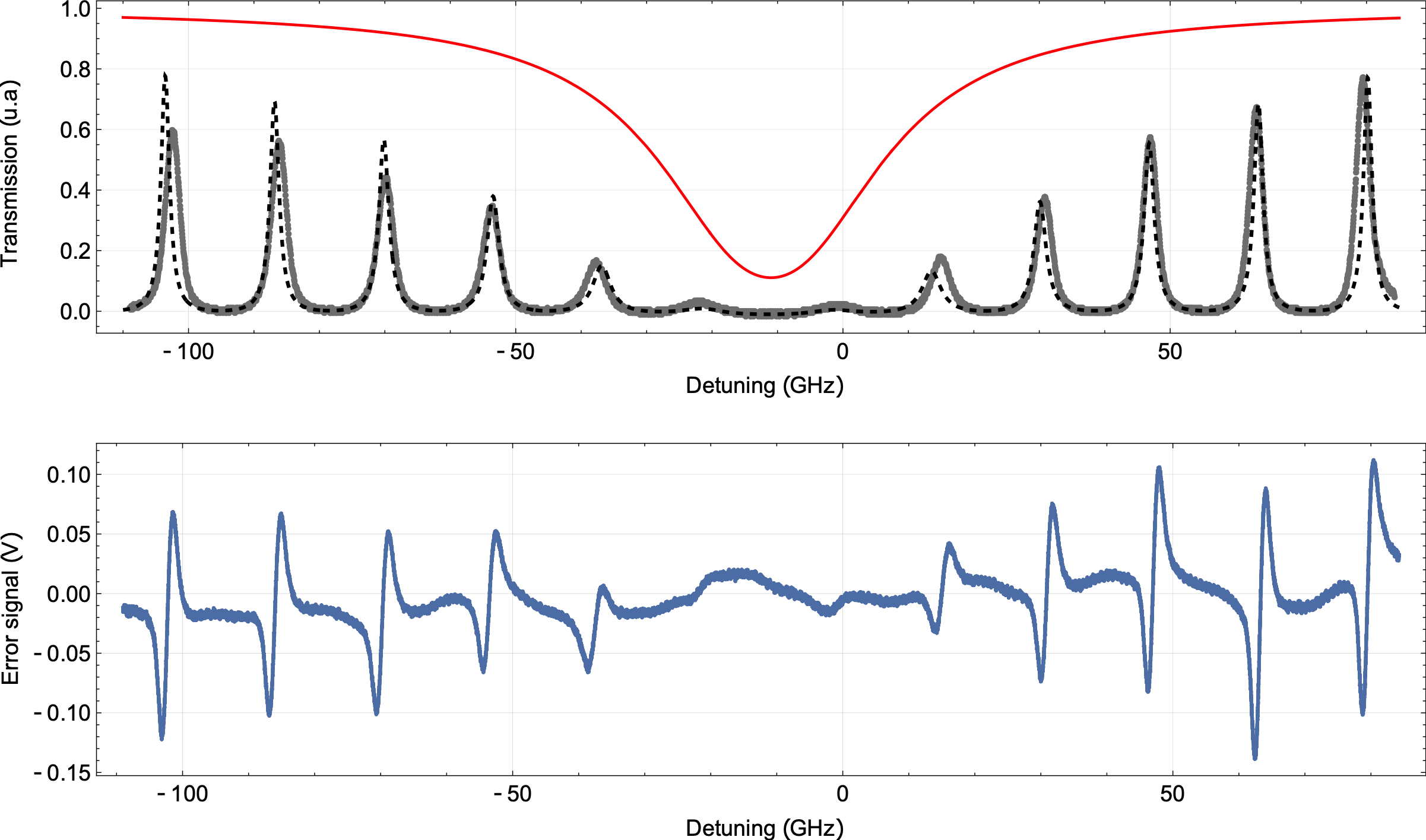} 
    \caption{Cavity transmission (top) and PDH error signal (bottom) near atomic resonance for a cavity containing a MEMS vapor cell. Detuning is measured with respect to the center of the \Dtwo{} transition of \textsuperscript{87}Rb, and is positive for blue detuning. The red curve shows the calculated single-pass vapor intensity transmission $|\tat|^2$. Black and blue solid lines show measured cavity transmission $|\tcav|^2$ and PDH error signal $\epsilon$, respectively. Dashed black line shows fitting of cavity transmission $|\tcav|^2$, with $\tcav$ given by ~\autoref{eq:CavityTAmpl} and  atomic medium single-pass transmission amplitude $t_A$ given by ~\autoref{Eq:ta}. For the fitting, the fixed parameters are the number density $n_A =  2.93\times10^{13}$ atoms/$\mathrm{cm^3}$, the mirrors reflectivities given in \autoref{Sec:Experimental system} and measured cell transmission amplitude $|\twin|^2=0.93$. Free parameters are  free spectral range $\FSR$, resonance shift $\Phi$, and an overall scale factor. }
    \label{fig:CavityTransmissionWithAtoms}
\end{figure*}

The Pound-Drever-Hall (PDH) method of cavity locking \cite{PoundRSI2004, DreverAPB1983, Black_PDH} is widely used in time-frequency metrology, and is analogous to frequency modulation spectroscopy techniques \cite{BjorklundAPB1983} used for high-resolution spectroscopy. In PDH, illustrated in \autoref{fig:PDHExpSetup}, a continuous-wave laser is phase modulated to produce sidebands above and below the carrier frequency. Upon reflection from the cavity, the carrier and side-bands acquire different, detuning-dependent phases, which converts some of the phase modulation to amplitude modulation. The resulting modulated power is directly detected with a fast photodetector and demodulated to obtain an error signal that, near resonance, is linear in the detuning, as seen in  \autoref{fig:CavityTransmissionWithAtoms}. The PDH method can be shot-noise limited, requires only a single optical polarization, and requires optical access to the cavity from only one side, all of which are attractive features for sensor miniaturization. Signal generation in PDH is described in greater detail in  Appendix~\ref{sec:TheoryOfCavityTransmissionReflectionAbsorption} to Appendix~\ref{sec:PDHSignalGeneration}.

Using the PDH method with vapors presents challenges
not encountered when using PDH for, e.g., laser stabilization to empty reference cavities. Vapors, especially those housed with buffer gas to slow diffusion, have multi-\SI{}{\giga\hertz} optical linewidths.  Together with high atomic number densities, of order \SI{1e13}{\per\centi\meter\cubed} or \SI{1e14}{\per\centi\meter\cubed}, these imply a significant broadening of cavity spectral lines due to atomic absorption, even in miniaturized cells. This absorption is moreover a function of the spin state, as the atoms can be found in states that are relatively ``dark'' or ``bright'' for the probe polarization. While quantitative models for atomic vapor optical properties \cite{HapperJauWalker}, optical cavities containing dispersive media \cite{BornAndWolfBook2019} and PDH signal acquisition \cite{Black_PDH} are all well established, to date no work has brought these together.

Here we demonstrate continuous PDH readout of the atomic spin dynamics of a \textsuperscript{87}Rb vapor housed in a MEMS cell. We use planar mirrors with dichroic coatings to resonate probe light detuned by $\sim \SI{100}{\giga\hertz}$ from the \Dtwo{} line at \SI{780}{\nano\meter}, while transmitting co-propagating pump light resonant with the  \Done{} line at \SI{795}{\nano\meter}.  We present a theoretical model that includes the complex refractive index of optically-pumped, collisionally-shifted and -broadened atomic vapor, cavity transmission and reflection in the presence of the resulting circular birefringence and circular dichroism, and PDH signal generation in such a cavity. The theoretical model is applicable to both the slow-modulation or ``unresolved sideband'' regime and the fast-modulation or ``resolved sideband'' regime. Our experiment is limited to low cavity finesse due to atomic absorption and cell losses, which practically restricts us to the unresolved sideband regime. We find excellent agreement of the model with experiment in this regime. 

The article is organized as follows.  \Cref{Sec:Experimental system} presents the experimental system, \Cref{sec:PhysicsAndOpticsOfPDHSpinMeasurement} describes a model encompassing the  A) Cavity resonance shifts due to spin polarization and B) the resulting PDH error signal. \Cref{sec:PDHSignalCharacterization} illustrates the experimental acquisition of PDH signals and demonstrates spin-dependent cavity line shifts. \Cref{sec:OPandNACalibration} calibrates the atomic response to optical pumping. \Cref{sec:ModulationAndRawPhotosignal} shows experimental modulation of the spin polarization using amplitude-modulated optical pumping and the resulting raw photocurrent signal, including quantum noise limited spin detection. \Cref{sec:DemodulatedPDH} characterizes the demodulated signals and the observed sensitivity to spin polarization density. \Cref{sec:Comparison} compares PDH with other methods. \Cref{sec:Outlook} describes possible extensions of the technique to improve sensitivity. 
 
\section{Experimental system}
\label{Sec:Experimental system}
The experimental setup is illustrated in \autoref{fig:PDHExpSetup}. In the PDH probing scenario an electro-optic modulator (EOM) applies phase-modulation at angular frequency $\Omega$ to a continuous-wave probe laser of angular frequency $\omegac$, chosen to be off-resonance with respect to an atomic transition of interest. The resulting beam is circularly polarized and mode-matched to a cavity containing an atomic medium. For this circular polarization, the refractive index and thus the cavity resonance frequency $\omegares$ are functions of $\Pat$, the on-axis component of the mean electron spin polarization  For simplicity, here and below we assume that collisional broadening of the optical transition dominates over Doppler broadening and hyperfine splittings. In this scenario, 1) the spin variable that determines the refractive index will be the electron spin $\mathbf{S}$ and 2) the transition lineshape can be approximated as a Lorentzian. In scenarios for which collisional broadening does not dominate, the optical properties may be different.  Upon reflection, the carrier and sidebands acquire phase shifts that depend on their respective detunings from cavity resonance, converting optical phase modulation into a detuning-dependent, and thus $\Pat$-dependent, amplitude modulation. The reflected beam is collected with a 125-MHz bandwith photodetector and the photocurrent is demodulated at frequency $\Omega$ with phase $\varphi$ to obtain the quadrature amplitude $\epsilon$, referred to as the PDH error signal. For laser-cavity detunings smaller than the cavity linewidth $\Gamma$, i.e. $|\omegac - \omegares| < \Gamma$, $\epsilon$ is approximately linear in $\omegac - \omegares$ and thus in $\Pat$. In our specific implementation two collinear lasers are used to pump and probe the atoms. The pump laser is resonant with the $^{87}$Rb \Done{} line, whereas the probe laser is detuned from the \Dtwo{} line by angular frequency $\Delta$.  A MEMS vapor cell contains isotopically-enriched $^{87}$Rb vapor and \SI{1.3}{\amagat} of N\textsubscript{2} due to prior UV decomposition of rubidium azide \cite{KarlenOE2017}. The MEMS cell consists of a silicon wafer, etched to create two $\SI{4}{\milli\meter} \times \SI{4}{\milli\meter}$ chambers joined by ``micro strainer'' conduits that allow gas and vapor diffusion between the chambers. Two \SI{200}{\micro\meter}-thick Borofloat windows, attached by anodic bonding, seal the MEMS cell. The Si wafer thickness, hence the experimental atomic interaction length is $\Lat=\SI{1.5}{\milli\meter}$. 
The vapor cell is placed between two half-inch planar mirrors, whose cell-facing surfaces are coated to give  reflectivity $R_1^{\Dtwo}=\SI{98.5}{\percent}$ and $ R_2^{\Dtwo} = \SI{99}{\percent}$ for the probe light (near the \SI{780}{\nano\meter} $\Dtwo{}$ line) and transmittivity $T_1^{\Done}=T_2^{\Done} \approx 50\%$ for the pump light (at the \SI{795}{\nano\meter} \Done{} line), where $R_i=|r_i|^2$ and $T_i=1-R_i$, $i\in \{1,2\}$ and $r_i$  are reflection amplitude coefficients. The mirrors are mounted on high-stability non-magnetic mounts, giving a free space cavity length of about $\Lcav=9.2$ mm. The cell is heated by alternating-current Joule heating to a temperature of \SI{126.5}{\celsius}, to achieve a number density of  $n_A\approx 2.93\times10^{13}$ atoms/cm$^3$. The system comprising the cavity, the MEMS cell, and the oven is placed within one layer of $\mu$-metal magnetic shielding.  
To implement the described PDH detection scheme, the output of an unmodulated distributed Bragg reflector (DBR) probe laser is phase modulated at $\Omega=2\pi\times\SI{114}{\mega\hertz}$ using a fiberized electro-optical modulator (NIR-MPX800-LN-05) driven by an amplified voltage-controlled oscillator. 
Pump light is provided by a second DBR laser diode amplified by a tapered amplifier,  which can generate up to \SI{2}{\watt} of continuous-wave power at \SI{795}{\nano\meter}. This output is passed through an acousto-optic modulator (AOM), whose first order diffraction output is coupled into fiber. Switching off the AOM radio-frequency drive is used to block or unblock the pump beam. A $1\times 2$ single-mode fiber coupler centered at 780 nm with $\pm$ 15 nm bandwidth (Thorlabs PFC780A) combines orthogonally polarized pump and probe light into a single spatial mode, which is collimated and aligned with the cavity. As measured by a beam profiler, the beam appears gaussian, with  beam diameters ($1/e^2$ of intensity)  of \SI{1.16}{\milli\meter} and \SI{0.91}{\milli\meter} in the x and y directions, respectively. As shown in \autoref{fig:PDHExpSetup}, a dual wavelength wave-plate (DWP) rotates the probe polarization such that both collinear beams are horizontally polarized along the y-direction to give maximum transmission through a polarizing beamsplitter (PBS). A zero-order quarter wave-plate (QWP) converts both to circular polarization. A second DWP can be inserted to control the direction of the pump beam circular polarization without changing the probe's circular polarization. Probe light reflected from the cavity is directed by the same PBS to a 125-MHz bandwith photodetector PD$_1$, whose output current is demodulated by the PDH circuit to provide the error signal, which is digitized by a data acquisition card (DAQ), as illustrated in \autoref{fig:PDHExpSetup}. The reflected pump light is filtered out at the detection stage by a bandpass filter centered at \SI{780}{\nano\meter} with \SI{3}{\nano\meter} bandwidth and an optical density (OD) of 6 at \SI{795}{\nano\meter}.

\begin{figure*}[t]
    \centering
    \includegraphics[scale=0.27]{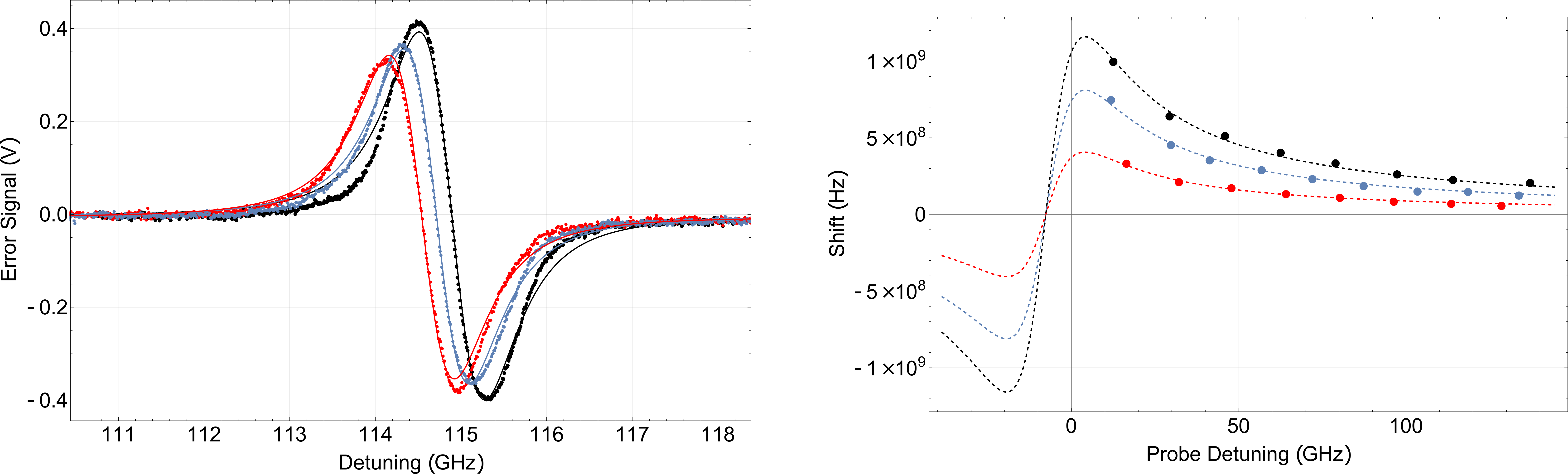}
    \caption{Response of PDH signal to optical pumping. (Left) PDH error signal $\epsilon$, obtained with  modulation frequency $\Omega= 2 \pi \times \SI{114}{\mega\hertz}$ and strong modulation depth $\beta=1.78$, in scans around a single cavity resonance. Detuning is measured with respect to the line-center of the atomic resonance of the transition \Dtwo{} of \textsuperscript{87}Rb. The probe light is  $\sigma_-$ polarized and detuned to the blue side of the atomic resonance. Blue dots show $\epsilon$ acquired with no optical pumping, red and black dots  show $\epsilon$ acquired with $\sigma_-$ and $\sigma_+$ optical pumping, respectively. The pump power is $\Ppump=\SI{22}{mW}$, which induces a degree of polarization of $P_z=0.65$ (calculated by using \autoref{eq:LinearFrequencyShift}). Continuous lines in the same colors show fits to the data with ~\autoref{Eq:ErrorSignal_generalBody}, with the resonance line center $\nu_p$ for polarization $p \in \{0, \sigma_\pm\}$, finesse $\mathcal{F}$ and system gain $K$ as free parameters.  Line shift due to $\sigma_+$ pumping is calculated as $\delta \nu_\mathrm{shift} = \nu_{\sigma_+} - \nu_0$. (Right) Line shift versus detuning from atomic resonance $\Delta/2\pi$ for different degrees of polarization. Red, blue and black points show $\delta \nu_\mathrm{shift}$ acquired by fitting scans, as just described, for optical pumping powers of \SI{5}{\milli\watt}, \SI{15}{\milli\watt} and \SI{55}{\milli\watt}, respectively. Each dot shows the average $\delta \nu_\mathrm{shift}$ of 20 pairs of fits. Dashed lines show fits of \autoref{eq:LinearFrequencyShift} to the measured shifts, and find $P_z = 0.29$, 0.57 and 0.82, from bottom to top. In these fits $P_z$ is the sole free parameter; fixed parameters  are $\Lat=\SI{1.5}{\milli\meter}$ (manufacturer's specification), $\Lcav=\SI{9.2}{\milli\meter}$ (from measured $\FSR$), $\Delta\nu_\mathrm{opt}\supDtwo=\SI{23.5}{\giga\hertz} =18.1\eta$ \cite{PhysRevA.56.4569}, (with measured N\textsubscript{2} density $\eta$=\SI{1.3}{\amagat}) and $n_A =  2.93\times10^{13}$ atoms/$\mathrm{cm^3}$.   
}  
    \label{fig:CombinedShifts}
\end{figure*}

\section{ PDH spin measurement}
\label{sec:PhysicsAndOpticsOfPDHSpinMeasurement}
We describe now the cavity response to an optically induced polarization. The optical response of the atomic vapor and the structure of the cavity resonances that determine the PDH signal are described in Appendices~\ref{sec:TheoryOfRefractiveIndex}
to \ref{sec:PDHSignalGeneration}.

\subsection{Cavity resonance shifts due to atomic spin polarization}
As derived in Appendix \ref{sec:TheoryOfCavityResonanceShifts}, the shift in the cavity resonance frequency due to polarization is given by: 
\begin{equation}
\delta \nu_\mathrm{} =
\frac{\pm \OD_0 c \PolEffect}{4 \pi \Lcav }
\frac{\nu}{\nu_\mathrm{at} }
\frac{(\Delta \nu_\mathrm{opt}\supDtwo/2)(\nu-\nu_\mathrm{at})}{(\nu-\nu_\mathrm{at})^2 + (\Delta \nu_\mathrm{opt}\supDtwo/2)^2}  P_z.
\label{eq:LinearFrequencyShift}
\end{equation}
where $\OD_0$ is the on-resonance optical depth of the unpolarized medium (Equation \ref{Eq:OpticalDepth}), $c$ is the speed of light, $\PolEffect = -2$ for the $\Done$ transition or $\PolEffect = 1$ for the $\Dtwo$ transition, $\nu$ is the resonance linear frequency, $\nu_\mathrm{at}$ and $\Delta \nu_\mathrm{opt}\supDtwo$ are, respectively, the line-center and FWHM of the atomic absorption resonance (linear frequency), and we have defined the degree of polarization $P_z \equiv 2 S_z$, equal to 1 for maximum spin polarization \cite{SeltzerThesis,Shah2009}.

\subsection{PDH error signal}

As described in Appendix~\ref{sec:PDHSignalGeneration}, the PDH signal after demodulation is
\begin{equation}
\label{Eq:ErrorSignal_generalBody}
    \epsilon =
K \sum_{k=-\infty}^{\infty} J_k(\beta)J_{k+1}(\beta)  2\mathrm{Re}\left[ \rcav^*(\omega_k)  \rcav(\omega_{k+1} )  e^{i \varphi}  \right],
\end{equation}
where $K$ is the overall gain of the system, including the laser power, photodetector responsivity, and the gains of the amplification and demodulation subsystems. $J_k(\beta)$ is the Bessel function of the first kind of order $k$, $\beta$ is the phase modulation amplitude, $\rcav$ is the frequency-dependent cavity reflection  amplitude given in \autoref{eq:CavityRAmpl}, $\omega_k \equiv \omegac + k \Omega$, where $\omegac$ is the probe carrier frequency, $\Omega$ is the modulation frequency and $\varphi$ is the demodulation phase.

\section{PDH signal characterization}
\label{sec:PDHSignalCharacterization}

For wide frequency scans over 200 GHz, as in \autoref{fig:CavityTransmissionWithAtoms}, we control the laser output frequency via the laser diode temperature. The frequency/temperature relation is monotonic but nonlinear, so for wide frequency scans, as in \autoref{fig:CavityTransmissionWithAtoms}, we use a fiber Mach-Zehnder interferometer with a \SI{41.28}{\mega\hertz} repeat period \cite{Kong_2015} to calibrate the frequency scale. We use as an absolute frequency reference either a wavemeter or the transmission of a spectroscopy cell containing \textsuperscript{87}Rb in vacuum.

The simultaneously-acquired transmission and PDH reflection signals are shown in \autoref{fig:CavityTransmissionWithAtoms}. Also shown is a fit of \autoref{eq:CavityTAmpl} to the transmission signal, which shows good agreement for the spectral positions and widths of the transmission peaks. In both the experiment and the model, the heights of the transmission peaks are greatly reduced near atomic resonance, due to intracavity absorption. The heights of the experimentally observed peaks show an asymmetry that is not present in the model. We attribute this to etalon effects produced by weak reflections from the cell windows, which are not included in the model. 

\newcommand{\MaxShiftOf}[1]{\delta\nu_\mathrm{max}{(#1)}}
\newcommand{\MaxShift}{\MaxShiftOf{\nu}}

\begin{figure}[t]
    \centering
     \includegraphics[width=\columnwidth]{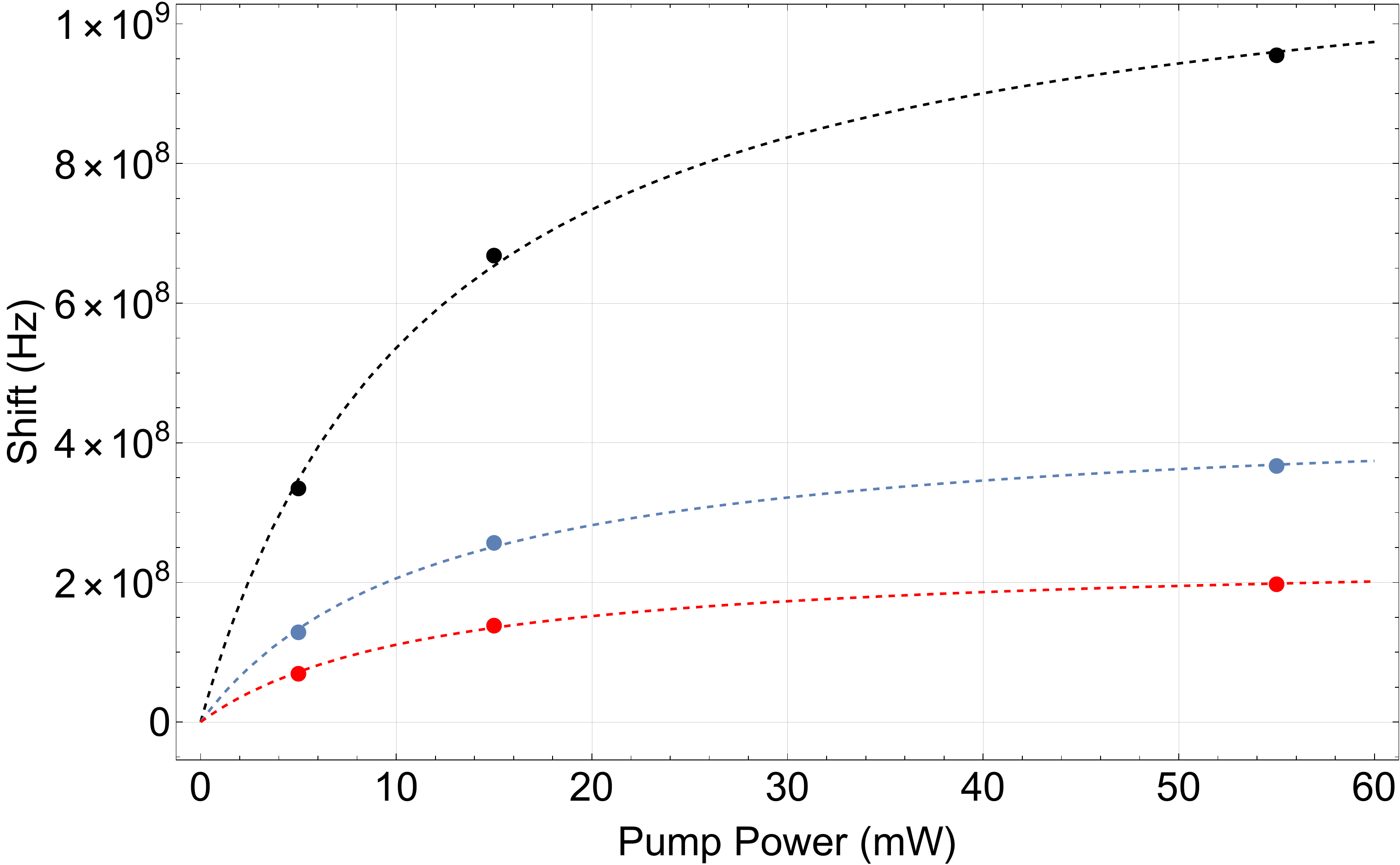}
    \caption{Shift of the error signal $\delta\nu_\mathrm{shift}$ versus pump power $\Popt$ for  probe detunings $\nu-\nu_\mathrm{at} = \SI{15}{\giga\hertz}$ (black), \SI{65}{\giga\hertz} (blue) and \SI{130}{\giga\hertz} (red). We take $\nu_\mathrm{at}$ to be the center of the \Dtwo{} transition of \textsuperscript{87}Rb. Points show $\delta\nu_\mathrm{shift}$ obtained by fitting pumped and unpumped single-resonance error spectra, as in \autoref{fig:CombinedShifts} (Left). Curves show fits with \autoref{eq:ParametrizedSaturationCurve}, with $\MaxShift$ and $\Psat$ as free parameters.  The fits find $\Psat=\SI{11.7}{\milli\watt}$, $\MaxShiftOf{\SI{15}{\giga\hertz}} =\SI{1.16}{\giga\hertz}$,  $\MaxShiftOf{\SI{65}{\giga\hertz}}= \SI{0.47}{\giga\hertz}$ and  $\MaxShiftOf{\SI{130}{\giga\hertz}}= \SI{0.24}{\giga\hertz}$}.  
    \label{fig:shift_pump_power}
\end{figure}

Focusing on a single resonance detuned by about \SI{115}{\giga\hertz}, \autoref{fig:CombinedShifts} (left panel) shows representative PDH error signals $\epsilon(\nu)$ for unpolarized atoms and for polarized atoms pumped with $\sigma_\pm$ circularly polarized light. The chosen detuning of  $\approx \SI{115}{\giga\hertz}$ reflects the trade-off between finesse, which is larger for larger detuning due to reduced atomic absorption, and spin-dependence of the resonant frequency, which is larger for smaller detuning. To acquire these spectra, we apply a \SI{500}{\hertz} triangle-wave modulation to the probe laser current, to scan the probe carrier frequency, while applying phase modulation as described in \autoref{Sec:Experimental system}. We synchronously apply a \SI{250}{\hertz} square-wave modulation to the RF driver of the AOM, so that the pump power is ``off'' for one scan of the resonance, and then ``on'' for the next. Combination and rotation of the QWP and  DWP is used to switch between $\sigma_+$ and $\sigma_-$ polarization of the pump beam.

Optical pumping induces a non-zero $S_z$ and thus modifies the refractive index $n$, both real and imaginary parts. Change in the real part $n'$ causes a shift of the resonance frequency, given by \autoref{eq:LinearFrequencyShift}, and a corresponding frequency shift of the error signal feature. Change of the imaginary part $n''$ alters the finesse of the resonance, and thus the peak-to-peak amplitude of the error signal. Both effects are clearly observed in ~\autoref{fig:CombinedShifts}. The error signals are fitted by using \autoref{Eq:ErrorSignal_generalBody} (details in caption of \autoref{fig:CombinedShifts}). 

~\autoref{fig:CombinedShifts} (right panel) shows the shift $\delta\nu\subres$ of the resonance frequency due to optical pumping, defined as the change in the zero-crossing of $\epsilon$ when pumped with $\sigma_+$ light, relative to the zero-crossing when unpumped. To measure this for a given probe detuning $\Delta$ and pump power $\Ppump$, we acquire 20 scans with pumping and 20 scans without pumping, fit them as described above, and compute the mean shift.  We acquired three different $\Ppump$ levels, each at  different values of $\Delta$. In order to get the different degrees of polarization for each $\Ppump$ level, we fit the data with \autoref{eq:LinearFrequencyShift}, the theoretical shift in resonance frequency, with only $P_z$ as free parameter.

\section{Optical pumping and atom number density calibration}
\label{sec:OPandNACalibration}
We describe the optical physics with a ``plane wave'' model of the optical pumping and cavity response. That is, we ignore the variation of optical intensity and resulting polarization with transverse position within the pump and probe beams. Optical pumping with $\sigma_+$ light, in competition with linear relaxation processes, produces an equilibrium spin polarization \cite{SeltzerThesis}
\begin{equation}
S_z =  \frac{\PolAdded\ROP}{2 \PolAdded\ROP+\RRel}
\label{Eq:Equilibrium_Polarization}
\end{equation}
where $\PolAdded$ is the average angular momentum added per absorbed photon and $\RRel$ is the spin relaxation rate.  The rate of optical pumping is $\ROP = \sigma_\mathrm{abs}(\nu) \phi_\mathrm{phot}$
\begin{eqnarray}
\ROP&=&\sigma_\mathrm{abs}(\nu) \phi_\mathrm{phot}
\nonumber \\ & = & 
r_e c \fosc^{\Done}\frac{\Delta\nu_\mathrm{opt}\supDone/2}{(\nu\subpump -\nu_\mathrm{at}\supDone)^2+(\Delta\nu_\mathrm{opt}\supDone/2)^2}\frac{\Ipump}{\hbar \omega}
\nonumber \\ & \stackrel{\mathrm{res}}{=} & 
\frac{r_e c \fosc^{\Done}}{\Delta\nu_\mathrm{opt}\supDone/2} \frac{\Ipump}{\hbar \omega},
\label{eq:ROPFromPPump}
\end{eqnarray}
where $\sigma_\mathrm{abs}$ is the absorption cross section and $\phi_\mathrm{phot}$ is the photon flux density, $\Ipump$ is the pump intensity, and the third line of \autoref{eq:ROPFromPPump} holds for resonant pumping $\nu = \nu_\mathrm{at}\supDone$, a condition we indicate with the label ``res''.

It is convenient to write \autoref{Eq:Equilibrium_Polarization} as
\begin{equation}
S_z \stackrel{\mathrm{res}}{=}  \frac{1}{2} \frac{\Ipump}{\Ipump+\Isat},
\label{Eq:Equilibrium_Polarization2}
\end{equation}

where the on-resonance saturation pump intensity is
\begin{equation}
   \Isat=\frac{ \Delta\nu_\mathrm{opt}\supDone  \hbar \omega}{ \PolAdded r_e c \fosc^{\Done}} \RRel.
    \label{Eq:DeltaNuMax}
\end{equation}

For quantitative modeling of the optical pumping, the spatial distribution of $S_z$ due to pump absorption \cite{Ito2016} and a consequent integral over the cell length should be considered \cite{Mizutani2014}. This description, however, is outside the scope of this current work.
Instead, we adopt the heuristic model 
\begin{eqnarray}
S_z &=& \frac{1}{2} \frac{ \Ppump}{\Ppump+\Psat}
\stackrel{\mathrm{ssr}}{\approx} 
\frac{1}{2} \frac{\Ppump}{\Psat},
\label{eq:ParametrizedSaturationCurveSz}
\end{eqnarray}
where $\Psat$ is the saturation pump power, and the approximation holds in the small signal regime (ssr) $\Ppump \ll \Psat$.  

\begin{figure*}[t]
\centering
\includegraphics[scale=0.87]{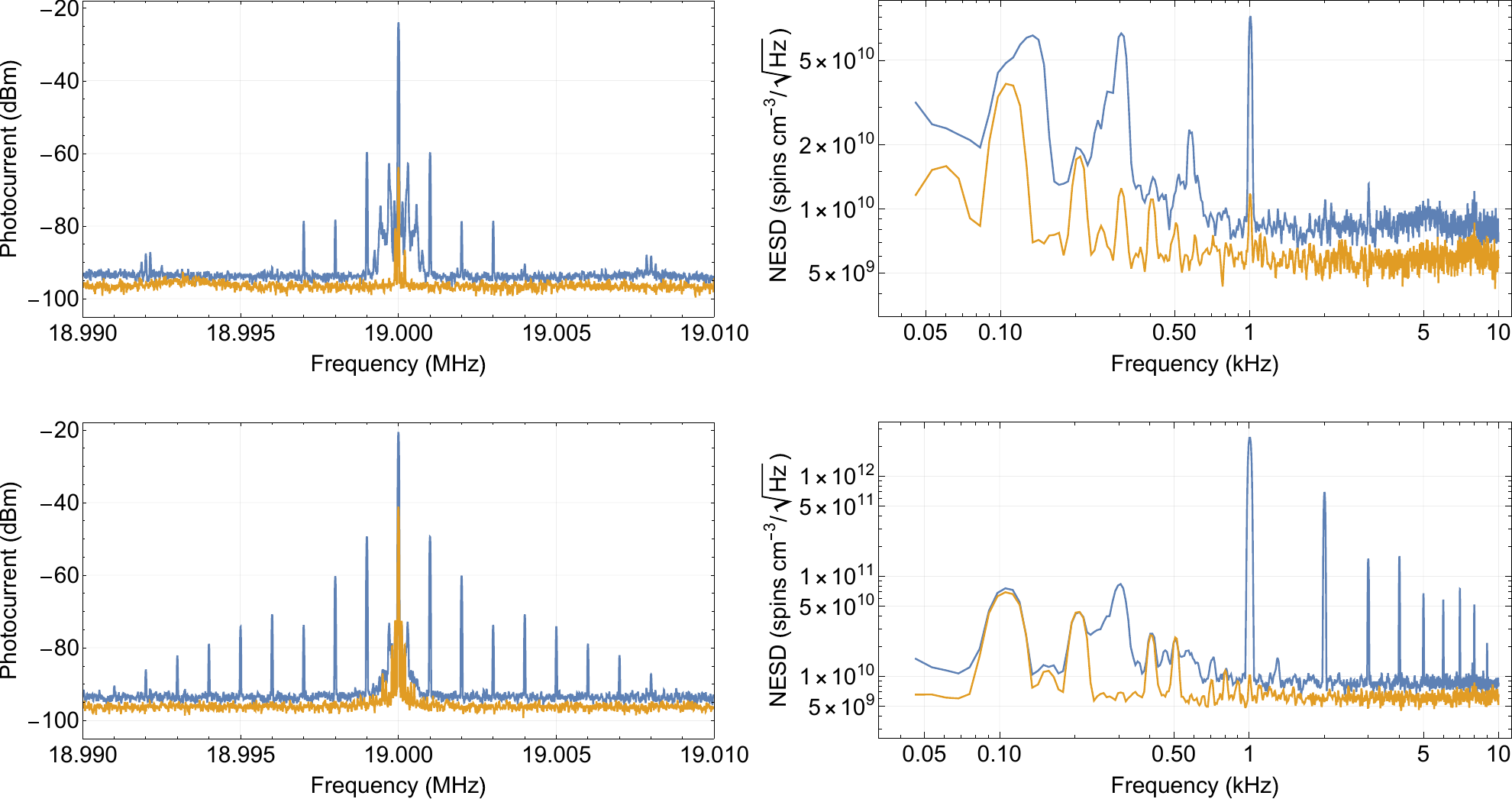}
\caption{ 
Signal and noise in PDH detection of $S_z$ modulation induced by optical pumping. Left graphs show spectrum analyzer traces of the detector photocurrent, expressed in electrical power units (dBm).  Right graphs show the spin sensitivity of the PDH error signal, i.e., of the photocurrent after demodulation, expressed as noise-equivalent spin density (NESD) with units of \SI{}{spins\per\centi\meter\cubed\per\sqrt\hertz} (see text for details of the calibration). Upper graphs were acquired in the small-signal regime with an optical pumping power of \SI{1}{\milli\watt} peak-to-peak sinusoidal modulation at \SI{1}{\kilo\hertz}. Lower graphs acquired in the large-signal regime with an optical pumping power of \SI{30}{\milli\watt} peak-to-peak sinusoidal modulation at \SI{1}{\kilo\hertz}. Orange curves, acquired with zero probe laser power, show detection and demodulation electronic noise background. Blue curves, acquired with \SI{2.8}{\milli\watt} of probe laser, detuned \SI{120}{\giga\hertz} (upper) or \SI{100}{\giga\hertz} (lower) to the blue of the center of the unshifted \Dtwo{} resonance, and with $\sigma_-$ polarization, include signal and optical noise. Spectra acquired with a digital spectrum analyzer with equal resolution bandwidth and video bandwidth $\RBW=\SI{30}{\hertz}$. 
}
\label{fig:noiseCharacterization}
\end{figure*}

Using \autoref{eq:LinearFrequencyShift}, this implies the resonance shift
\begin{eqnarray}
\delta\nu &=& \delta\nu_\mathrm{max}(\nu\subprobe) \frac{ \Ppump}{\Ppump+\Psat},
\label{eq:ParametrizedSaturationCurve}
\end{eqnarray}
where for $\sigma_\pm$ probe light 
\begin{equation}
\delta \nu_\mathrm{max}(\nu) =
\frac{\pm \OD_0 c g}{ 4\pi \Lcav }
\frac{\nu}{\nu_\mathrm{at} }
\frac{(\Delta \nu/2)(\nu-\nu_\mathrm{at})}{(\nu-\nu_\mathrm{at})^2 + (\Delta \nu/2)^2}
\label{eq:LinearFrequencyShift2}
\end{equation}
is the asymptotic frequency shift.   \autoref{fig:shift_pump_power}, shows fits of \autoref{eq:ParametrizedSaturationCurve} to data (the same results as in \autoref{fig:CombinedShifts} (right panel)), to find $\Psat = \SI{11.7}{\milli\watt}$ and $n_A =  2.93\times10^{13}$ atoms/$\mathrm{cm^3}$ (in $\OD_0$), with good agreement.

\section{Modulation of $S_z$ and raw photosignal}
\label{sec:ModulationAndRawPhotosignal}

To study the response of the cavity to changes in spin polarization, we apply phase modulation to the probe as described in \autoref{Sec:Experimental system}. Because our photodetector is not shot-noise limited at the \SI{114}{\mega\hertz} frequency used until now, for the remaining measurements we use a modulation frequency of \SI{19}{\mega\hertz}. The signal from \PDone{} (See \autoref{fig:PDHExpSetup}) is split into dc and rf components with a bias-tee. The rf component is amplified and further split with a \SI{3}{\decibel} splitter, one output of which goes to  a spectrum analyzer (SA), the other of which goes to the demodulation system (mixer and low-pass filter). Based on the demodulated signal, the
probe carrier frequency $\omega_0$ is manually set to the center of the resonance and during the acquisition remains within the central region, i.e., the region of negative slope in \autoref{fig:CombinedShifts} (left). Representative spectra are shown in \autoref{fig:noiseCharacterization} (left panels).

We apply a sinusoidally-modulated pump power 
\begin{equation}
\Ppump = \Ppump\supzero + \frac{1}{2} \Ppump\supPP \cos \OmegaPump t,
\end{equation}
where $\Ppump\supPP$ is the peak-to-peak variation of the pump power and $\OmegaPump/2\pi \approx \SI{1}{\kilo\hertz}$ is the modulation frequency. This induces a sinusoidal shift in the PDH line center. This, in turn, produces a periodic modulation at frequency $\OmegaPump$ of the  amplitude modulation at $\Omega$ seen by \PDone. This manifests as signal sidebands at $\Omega \pm \OmegaPump$, and  distortion sidebands at $\Omega \pm j \OmegaPump$, $j = 2, 3, \dots$.  We study two signal regimes: with 
$\Ppump\supPP = \SI{1}{\milli\watt}$ of resonant pump power, we access the small-signal regime, in which the root-mean-squared (RMS) line-shift $\delta\nu_\mathrm{shift}\supRMS \ll \Gammacav$, and the distortion sidebands are nearly \SI{20}{dBm} below the signal sidebands.  With $\Ppump\supPP = \SI{30}{\milli\watt}$, we access the large-signal regime, with $\delta\nu_\mathrm{shift}\supRMS \sim \Gammacav/2$. 
With this $\Ppump\supPP$, the distortion sidebands are about \SI{10}{dBm} below the signal sidebands.  

These spectra also show a strong peak at the probe phase modulation frequency $\Omega = 2\pi \times \SI{19}{\mega\hertz}$, which results from drift of $\omega_0$ away from line center. Noise in the band $\Omega- 2\pi \times \SI{700}{\hertz}$ to $\Omega+ 2\pi \times \SI{700}{\hertz}$ is comparable to or weaker than the applied signal. The noise in this frequency band can be attributed to small and slow shifts in the probe laser frequency, e.g., those due to laser temperature changes. 

Other frequency bands show a flat noise spectrum about \SI{2}{\decibel} above the electronic noise level of the acquisition, obtained with $P\subprobe=0$. We interpret this \SI{2}{\decibel} contribution as quantum noise: the sum of photon shot noise and the broadband component of atomic spin projection noise \cite{MouloudakisPRA2022}. We note that the noise background does not rise in going from the small-signal to large-signal regimes. This indicates a negligible level of technical noise associated with the optical pumping of the spins and with the increased signal strength.

\section{Demodulated PDH signal and sensitivity}
\label{sec:DemodulatedPDH}

In \autoref{fig:noiseCharacterization} (right panels), we show spectra of the demodulated photocurrent, i.e., the PDH error signal $\epsilon$, acquired under the same conditions as described in  \autoref{sec:ModulationAndRawPhotosignal}. As expected for the demodulation process, the spectrum strongly resembles that of \autoref{fig:noiseCharacterization} (left panels), downshifted by the modulation frequency $\Omega$.

A narrow peak at \SI{1}{\kilo\hertz} is the signal, due to applied pump modulation. Noise in the band below $\approx \SI{700}{\hertz}$ shows peaks at the \SI{50}{\hertz} mains frequency and harmonics. Outside this band, the noise is spectrally flat, and once again we interpret this as the combined photon shot noise and spin projection noise. Again, we note the near-absence of detectable technical noise associated with optical pumping and strong signal, with the possible exception of a small peak near \SI{1.2}{\kilo\hertz}.

We identify the  sensitivity of the measurement, i.e., equivalent spin noise spectral density, as follows. Considering the trace taken in the small-signal regime, for which we can use \autoref{Eq:Equilibrium_Polarization} to convert from $\Ppump$ to $S_z$, we note that the applied monochromatic pump power modulation, with mean squared variation $(\Ppump\supPP)^2/8$,  produces by \autoref{eq:ParametrizedSaturationCurveSz} a mean square variation in $\nat S_z$ of 
\begin{eqnarray}
\mathcal{P}_{\nat S_z} &=& 
\frac{n_A^2}{8}
\left( \frac{ P_\mathrm{pump}\supPP}{2 \Psat} 
\right)^2.
\end{eqnarray}
This single-frequency ``signal-like'' feature is represented in the SA's spectrum by a peak of height $\PSARef \equiv \PSA(\nuPump)$ with units of 
electrical power. The conversion from pump power to spin density signal power is thus 
$\mathcal{P}_{\nat S_z} / \PSARef$. Meanwhile, the broad ``noise-like'' features of the same spectrum must be interpreted as an electrical power  spectral density $\PSA(\nu)/\RBW$, or after conversion to spin units, the spin-noise power spectral density
\begin{equation}
\mathcal{S}_{\nat S_z}(\nu\subsig) = \frac{P_\mathrm{SA}(\nu\subsig) }{\PSARef \RBW} \frac{n_A^2}{8} \left(\frac{\Ppump\supPP}{2\Psat} \right)^2,
\end{equation}
where all $P$ quantities have units of electrical power ($P_\mathrm{SA}(\nu\subsig)$ and $\PSARef$) or optical power ($\Ppump\supPP$ and $\Psat$).  $\mathcal{S}_{\nat S_z}(\nu\subsig)$ is a power spectral density with units of \SI{}{spins^2\per\centi\meter^6\per\hertz}.  In \autoref{fig:noiseCharacterization} (right), we graph the square root of this quantity, the amplitude spectral density (ASD), with units of \SI{}{spins\per\centi\meter\cubed\per\sqrt\hertz}. In analogy to the {noise-equivalent power} that describes detector sensitivity, $\mathcal{S}_{\nat S_z}^{1/2}(\nu\subsig)$ is the {noise-equivalent spin density} of the measurement.
\newcommand{\SNR}{\mathrm{SNR(\phi)}}
\newcommand{\MSE}{\mathrm{MSE(\phi)}}

\section{Comparison to other methods}
\label{sec:Comparison}
As described in \autoref{sec:Introduction}, a variety of optical enhancement methods, including  multi-pass and cavity-enhancement, have been used for sensing and metrology applications that require sensitive non-destructive measurement of atomic polarization. While it is clear that optical enhancement methods improve  relevant figures of merit for these applications, e.g., magnetic sensitivity for magnetometers \cite{ShengPRL2013, Crepaz2015}, scattering for atomic state measurement \cite{Mazzinghi2021}, or short-term stability for atomic clocks \cite{LodewyckPRA2009}, to our knowledge there is no established figure of merit that allows comparison across applications. We can nonetheless note some commonalities. In multi-pass methods with fixed medium length, the signal, e.g., an accumulated phase or Faraday rotation angle, grows linearly with the number of passes. Similarly, in cavity enhancement, again with fixed cavity length and thus fixed FSR, the slope $d(\arg(\rcav))/d\omega$ of the reflected phase is inversely proportional to the cavity linewidth, and thus proportional to the finesse.  We can thus identify the number of passes and the finesse as roughly-comparable measures for signal enhancement. Number of passes range from 10s in magnetometry \cite{ShengPRL2013} to 100s in trace-gas spectroscopy \cite{DasAPB2011}, while finesses range from $\sim 10$ in $\sim \SI{10}{\centi\meter}$ cavities containing vapor cells \cite{Crepaz2015, Mazzinghi2021} to $\sim 10^5$ for cold atom cavities in vacuum \cite{Hosten2016}.  The finesse in this experiment depends on the intracavity atomic absorption and thus the detuning. For a detuning of $2\pi \times \SI{115}{\mega\hertz}$ as in \autoref{fig:CombinedShifts}, we have $\mathcal{F} \approx $ 18, similar to values obtained with larger vapor cells.

\section{Outlook}
\label{sec:Outlook}

We note a few possible improvements. The current system operates in the unresolved sideband regime, with $\Omega \ll \Gamma$. In this regime, the signal strength is proportional to $\Omega$. A better sensitivity can thus be expected simply by operating at higher $\Omega$. To do so with shot-noise-limited performance would require a different detector, one that is shot-noise limited at similar power levels and higher frequencies. We note that using specific detector geometries, detectors with shot-noise-limited bandwidths of \SI{9}{\giga\hertz} \cite{TaskerNP2021} have been demonstrated, with potential to reach still much higher frequencies \cite{LischkeOE15}. 

Our system is implemented with a microfabricated vapor cell inside a cavity formed by independent mirrors. Losses on the cell windows limit the finesse of the resonant cavity in this geometry. A vapor cell in which the internal surfaces were coated would have a finesse limited in principle only by the reflectivity of the cavity mirrors, which would double as interior surfaces of the vapor cell. We note that \textsuperscript{87}Rb MEMS cells with (non-optical) internal coatings \cite{KarlenOE2017} have been demonstrated.

\section{Conclusions}

We have implemented optically-resonated readout of the spin polarization of an atomic vapor in a microfabricated vapor cell. We use a planar cavity, resonant for the \Dtwo{} line of \textsuperscript{87}Rb, and Pound-Drever-Hall detection of cavity line shifts produced on circularly-polarized probe light, to detect the electron spin polarization $S_z$ of a dense vapor \textsuperscript{87}Rb in \SI{1.3}{\amagat} of N\textsubscript{2} buffer gas. We present a model for the resulting signals, combining theory of the vector polarizability of alkali vapors, resonant cavity response, and Pound-Drever-Hall signal generation. We find good agreement of the model with experimental observations, and demonstrate a quantum-noise-limited sensitivity to spin density of $\approx \SI{9e9}{spins\per\centi\meter\cubed\per\sqrt\hertz}$. {The technique has potential to provide high efficiency readout for miniaturized atomic vapor sensing and metrology instruments, in the same way that cavity-enhanced readout has been used with cold atom instruments.} 

\section{Acknowledgments}
We thank Jakob Reichel for insights about cavity enhancement, Kostas Mouloudakis, Michael Tayler and Aleksandra Sierant for laboratory assistance and helpful discussions, and Jacques Haesler, Sylvain Karlen and Thomas Overstolz of the Centre Suisse d'Electronique et de Microtechnique SA (CSEM) in Neuch{\^a}tel (Switzerland) for providing MEMS vapor cells. This work was supported by 
European Commission projects MACQSIMAL (820393), OPMMEG (101099379) and QUANTIFY (101135931) and NextGenerationEU (PRTR-C17.I1), Spanish Ministry of Science (MCIN) project SAPONARIA (PID2021-123813NB-I00) and ``Severo Ochoa'' Center of Excellence CEX2019-000910-S, Departament de Recerca i Universitats de la Generalitat de Catalunya grant No. 2021 SGR 01453;  Fundaci\'{o} Privada Cellex; Fundaci\'{o} Mir-Puig; 
MHR acknowledges support from Ayuda PRE2021-098880 financiada por MCIN/AEI/ 10.13039/501100011033 y por el FSE+. VGL acknowledges financial support from European Union NextGenerationEU (PNRR MUR project PE0000023 – NQSTI) and from the Italian Ministry of University and Research (MUR) project "Budget MIUR - Dipartimenti di Eccellenza 2023 - 2027" (Law 232, 11 December 2016) - Quantum Sensing and Modelling for One-Health (QuaSiModO). HM acknowledges financial support from the European Union’s Horizon Europe research and innovation programme
under the Marie Sklodowska-Curie grant agreement No. 101081441. YM
acknowledges the support from China Scholarship Council (202206280171).

VGL and MWM jointly supervised this work. 

\appendix

\section{Atomic vapor refractive index}
\label{sec:TheoryOfRefractiveIndex}

For low density media such as vapors, local-field effects can be neglected and the electric susceptibility calculated as $\chi = \nat \alpha$ \cite{YarivBook1989}, where $\nat$ is the atomic number density and $\alpha$ is the atomic polarizability. In general, $\chi$ and $\alpha$ are complex-valued rank-two tensors, as required to describe circular dichroism and circular birefringence.  Also in these conditions, the refractive index is given, in cgs units, by $n = \sqrt{1 +4 \pi \nat \alpha} \approx 1 + 2\pi \nat \alpha$ or \begin{eqnarray}
\label{eq:IndexSusceptibilityRelation} 
(n -1) & \approx& 2\pi \nat \alpha.
\end{eqnarray}

The below calculations concern the linear-optical susceptibility and polarizability, and assume the optical transitions are not saturated by probe or pumping light. A general condition for the validity of this assumption is the following \cite{YarivBook1989}:
\begin{eqnarray}
\label{eq:rabicond2}
\Omega_\mathrm{opt} \ll \frac{1+\Delta^2 \Gamma^{-2}_2}{4 \Gamma^{-1}_1 \Gamma^{-1}_2},
\end{eqnarray}
where $\Omega_\mathrm{opt} = |E \cdot d_{ba}|/\hbar$ is the optical Rabi frequency,  $d_{ba}$ is the optical transition electric dipole moment between ground state $|a\rangle$, and excited state $|b\rangle$, and $\Gamma_1$ is the corresponding longitudinal coherence relaxation rate. \autoref{eq:rabicond2} holds in all the experimental conditions we present in this work.

\label{sec:TensorOperatorRefractiveIndex}
The calculation of the atom's state-dependent electric polarizability tensor $\alpha$ is extensively discussed in  the  optical pumping  literature, for example in \cite{HapperMathurPR1967, HapperRMP1972, HapperJauWalker}. The ac-Stark shift of the ground state, obtained by taking the electric dipole interaction in second order and by adiabatic elimination of the excited state, is given by \cite[Eq.~6.73]{HapperJauWalker} 
\begin{equation}
\delta H = -\mathbf{E} \cdot \alpha \cdot \mathbf{E}^*,  
\end{equation} 
where $\mathbf{E} = |\mathbf{E}| \mathbf{e}$ is the complex vector amplitude of the optical electric field. We work with quantization axis $z$, so the $\sigma_\pm$ and $\pi$ polarizations are described by $\mathbf{e}_\pm \equiv (\hat{\mathbf{x}} \pm i \hat{\mathbf{y}})/\sqrt{2}$ and $\mathbf{e}_0 \equiv \hat{\mathbf{z}}$, respectively. $\alpha$ is in general a second-rank tensor, with irreducible scalar, vector (``gyrotropic''), and tensor (``birefringent'') parts.  For our experimental conditions, the collisional broadening is much larger than the excited state hyperfine splitting and the Doppler broadening. Consequently, the optical transition can be treated as homogeneously broadened, and the tensor part of the susceptibility can be neglected, leaving only the scalar and vector parts. In these conditions, the susceptibility in Cartesian coordinates is \cite[Eqs.~5.41, 6.73, 6.75]{HapperJauWalker}
\begin{equation}
\alpha_{ij} = {\alpha_0} \left( \frac{1}{2}\delta_{ij} + \frac{2(-1)^{J-1/2}}{[J]} i \epsilon_{ijk}S_k \right),
\end{equation}
where $\delta_{ij}$ is the Kronecker symbol and $\epsilon_{ijk}$ is the Levi-Civita symbol. Summation over repeated indexes is implied, and \cite[Eq.~5.45]{HapperJauWalker}
\begin{equation}
\alpha_0 = \frac{r_e c^2 \fosc}{\omegaA} \frac{-1}{\Delta + i \Gamma_2},
\end{equation}
where $\fosc$ is the relevant oscillator strength. 
Within the planar resonator of the optical cavity, the optical field is to a good approximation a plane standing wave, propagating along the $\pm z$ directions. For this reason we neglect the longitudinal, i.e.,  $\mathbf{e}_0= \hat{\mathbf{z}}$, component of the field. The relevant portion of $\alpha_{ij}$,  in Cartesian components ($\hat{\mathbf{x}}$, $\hat{\mathbf{y}}$), is then
\begin{equation}
\alpha_{ij} = {\alpha_0} \left[ \frac{1}{2} \left( 
\begin{array}{cc}
     1 & 0 \\
     0 & 1
\end{array} \right) + 
\frac{2(-1)^{J-1/2} S_z}{[J]} 
\left( 
\begin{array}{cc}
     0 & i \\
     -i & 0
\end{array} \right) \right],
\end{equation}
which has eigenvectors $\mathbf{e}_\pm$, corresponding to $\sigma_\pm$ polarization. For the \Done{} ($J=1/2$) and \Dtwo{} ($J=3/2$) transitions, the corresponding eigenvalues are
\begin{eqnarray}
\alpha_{\pm}^{\Done} & = & \alpha_0\left( \frac{1}{2} \mp S_z\right)  \\
\alpha_{\pm}^{\Dtwo} & = & \alpha_0\left( \frac{1}{2} \pm \frac{1}{2} S_z\right).
\end{eqnarray} 
From this we see that other components of $\mathbf{S}$ do not contribute to the polarizability.

Using \autoref{eq:IndexSusceptibilityRelation},  $\sigma_\pm$ light on the \Dtwo{} and \Done{} transitions experiences  
\begin{eqnarray}
\label{eq:IndexForDOne1} 
n_\pm^{\Done} -1 & \approx& \frac{\pi \nat r_e c^2 \fosc^{\Done}}{\omegaA^{\Done}} \frac{-1}{\Delta + i \Gamma_2}
\left( 1 \mp 2 S_z\right) \\
\label{eq:IndexForDTwo1} 
n_\pm^{\Dtwo} -1 & \approx& \frac{\pi \nat r_e c^2 \fosc^{\Dtwo}}{\omegaA^{\Dtwo}} \frac{-1}{\Delta + i \Gamma_2}
\left( 1 \pm S_z\right), \hspace{6mm}
\end{eqnarray}
where $\fosc^{\Done} \approx 0.3423$ and $\fosc^{\Dtwo} \approx 0.6958$ are the transition oscillator strengths.

\section{Spin-dependent medium refractive index}
\label{sec:TheoryOfSpinDependentRefractiveIndex}

As described in Appendix~\ref{sec:TheoryOfRefractiveIndex}, in conditions relevant to many miniaturized sensors, i.e., realistic vapor density, high buffer gas pressure, and optical powers that do not saturate the optical transition \footnote{The model does apply for optical powers that saturate optical pumping.}, the refractive index for light near the $\Done$ or $\Dtwo$ transition, with $\sigma_\pm$ circular polarization (quantization axis is along $z$) is
\begin{eqnarray}
\label{eq:IndexFromS} 
n_\pm -1 & \approx& \frac{\pi \nat r_e c^2 \fosc}{\omegaA} \frac{-1}{\Delta + i \Gamma_2}
\left( 1 \pm \PolEffect S_z \right) 
\end{eqnarray}
where $n_A$ is the atomic number density, $r_e$ is the classical electron radius, $c$ is the speed of light, $\fosc$ is the transition oscillator strength ($\fosc^{\Done} \approx 0.3423$ and  $\fosc^{\Dtwo} \approx 0.6958$ in \textsuperscript{87}Rb), $\omegaA$ is the transition angular frequency,  $\Delta \equiv \omega - \omegaA$ is the detuning, $\omega$ is the optical angular frequency, $\Gamma_2$ is the transverse relaxation rate of the optical transition, and thus the (angular frequency) half-width at half-maximum  of the collisionally-broadened absorption spectrum, $\PolEffect = -2$ for the $\Done$ transition or $\PolEffect = 1$ for the $\Dtwo$ transition, and $S_z$ is the on-axis spin polarization.  For brevity, we define $n' \equiv \mathrm{Re}[n]$ and $n'' \equiv \mathrm{Im}[n]$.  

Upon passing through a cell of length $\Lat$, the transmission amplitude attributable to the atoms (as opposed to the distance travelled or atom-independent losses) is
\begin{eqnarray}
t_A &=& \exp[i \frac{(n-1) \omega}{c} \Lat]. 
\label{Eq:ta}
\end{eqnarray}
We note that $n$ is linear in $S_z$, that the real and imaginary parts of $(n-1)$ are affected by $S_z$ in the same proportion, and that for both $\Done$ and $\Dtwo$ the sign of $dn/dS_z$ reverses with a change in circular polarization. 
The on-resonance optical depth of the unpolarized medium is 
\begin{equation}
\label{Eq:OpticalDepth}
\OD_0 = \frac{n'' \omegaA \Lat}{c} = \frac{\pi \nat r_e c \fosc \Lat}{\Gamma_2} 
\end{equation}
and it is convenient to write \autoref{Eq:ta} as
\begin{eqnarray}
\label{eq:IndexFromSandOD} 
\tat & = & \exp \left[ i  \frac{ -\Gamma_2}{\Delta + i \Gamma_2} (1 \pm \PolEffect S_z) \OD_0 \right].
\end{eqnarray}

\section{Cavity transmission, reflection and absorption}
\label{sec:TheoryOfCavityTransmissionReflectionAbsorption}

The cavity is formed by mirrors $\mathrm{M}_{i}$, $i\in \{1,2\}$, with $1$ being the mirror via which laser power enters, and $2$ being the other mirror, with amplitude transmission and reflection coefficients $t_i$ and $r_i$,  respectively. We adopt the convention that all mirror reflection and transmission coefficients are positive real, with the exception of reflection of a beam approaching from outside the cavity, in which case the reflection amplitude is negative real. We assume lossless mirrors, and thus $r_i^2 + t_i^2 = 1$ (equivalent losses will be incorporated elsewhere in the model). The cavity contains a vapor cell of interior length $\Lat$ and single-pass transmission amplitude $\tcell = \twin^2 \tat  \exp[i \Lat \omega/c]$, where $\twin$ is the amplitude transmission of a single cell window. Due to scattering losses, $|\twin|^2$ may be less than unity.  $\tat$, defined in \autoref{Eq:ta}, is the atomic contribution. The optical path length between the mirrors, not including the atomic contribution to $n$, is $\Lcav$, so that for an empty cell, or in the limit of large detuning from atomic resonance, the free spectral range (FSR, a linear frequency) is $\FSR \equiv c/2\Lcav$.

The frequency-dependent cavity transmission amplitude can be calculated by summing the transmission amplitude for all trajectories  \cite{BornAndWolfBook2019}, as 
\begin{eqnarray}
\tcav &=&  t_1 \tcell  t_2  \exp[i \frac{(\Lcav-\Lat) \omega}{c}]
\nonumber \\ & &  \times 
\sum_{n=0}^{\infty} \left(r_2 r_1  \tcell^2   \exp[i \frac{2(\Lcav-\Lat) \omega}{c}] \right)^n  \hspace{1mm} \nonumber \\ & =& 
\frac{t_1 t_2 |\twin^2|  \tat  \exp[i(\omega/\FSR + \Phi)/2] } {1- r_1 r_2  |\twin^4|  \tat^2  \exp[i( \omega/\FSR + \Phi)]}
\label{eq:CavityTAmpl}
\end{eqnarray}
where $\Phi \equiv \arg(\twin^4)$. The cavity transmission probability is thus
\begin{eqnarray}
|\tcav|^2 &=& \left|
\frac{t_1 t_2 \twin^2 \tat } {1- r_1 r_2  |\twin^4|  \tat^2  \exp[i( \omega/\FSR + \Phi)]}\right|^2. \hspace{6mm}
\label{eq:CavityTProb}
\end{eqnarray}

In calculating the above we have neglected etalon effects from reflections from the interfaces of the cell windows \footnote{These amplitudes are in practice much smaller than the near-unity reflection amplitude of the cavity mirrors, and can be made still smaller through anti-reflection coating. Their inclusion significantly complicates the resulting expressions without changing qualitatively the results.}. The etalon effects can explain for the asymmetry that is observed in the experimentally measured transmission peaks shown in Figure \ref{fig:CavityTransmissionWithAtoms} . 

The reflection amplitude is computed in the same way, to find
\begin{eqnarray}
    \rcav &=&  -r_1 + t_1^2 r_2 \tcell^2  \exp[i 2 (\Lcav - \Lat) \omega/c]   
\nonumber \\ & & \times \sum_{n=0}^{\infty} \left(r_1 \tcell^2  r_2 \exp[i 2 (\Lcav - \Lat) \omega/c]  \right)^n 
\nonumber \\ &=& 
\frac{-r_1 +  r_2 |\twin^4| \tat^2   \exp[i (\omega /\FSR + \Phi)]}{1- r_1 r_2 |\twin^4| \tat^2  \exp[i (\omega /\FSR + \Phi)]}.  
\label{eq:CavityRAmpl}
\end{eqnarray}
Resonance occurs when $\mathrm{mod}_{2\pi}(2 \arg \tat + \omega/\FSR + \Phi) = 0$.  The cavity resonances have finesse $\mathcal{F}\approx\pi/(1-|r_1 r_2 \twin^4 \tat^2 |)$ and full-width at half-maximum (FWHM) linewidth $\Gammacav \equiv \FSR/\cal{F}$, which will depend on detuning from the atomic resonance.  

\section{Cavity resonance shifts due to atomic spin polarization}

\label{sec:TheoryOfCavityResonanceShifts}
From Eqs.~\ref{Eq:ta}, \ref{eq:CavityTProb} and \ref{eq:CavityRAmpl}, we see that the dependence of $\rcav$ and $|\tcav|^2$ on probe frequency $\omega$ and spin polarization $S_z$, through the refractive index $n$, is contained in the factor
\begin{equation}
\tat^2 \exp\left[i\left(\frac{\omega}{\FSR}+ \Phi\right)\right]  =  \exp[-2\frac{n''\omega}{c} \Lat]  \exp[i \phi(\omega)] 
\end{equation}
where the net intra-cavity round-trip phase is
\begin{equation}
\phi(\omega) \equiv  2\frac{(n'-1)\omega}{c} \Lat+   \frac{\omega}{\FSR} +\Phi.
\label{eq:phidef}
\end{equation}

In terms of $\phi$, the optical resonance defined after \autoref{eq:CavityRAmpl} occurs when $\mathrm{mod}_{2\pi}\phi(\omega)  = 0$. Maintaining resonance by holding $\phi$ constant, and using \autoref{eq:IndexFromSandOD}
\begin{eqnarray}
\left( \frac{\partial \omega}{\partial S_z} \right)_{\phi} &=& -\left( \frac{\partial \phi}{\partial \omega} \right)^{-1}\left( \frac{\partial \phi}{\partial S_z} \right) 
\nonumber \\ 
& = & 
\frac{\pm \omega  }{(n'-1)\Lat + \Lcav} 
\frac{\OD_0  c  \PolEffect }{ \omegaA }
\frac{\Delta \Gamma_2}{\Delta^2 + \Gamma_2^2}. \hspace{6mm}
\nonumber \\ 
& \approx & 
\frac{\pm \omega }{\Lcav} 
\frac{\OD_0  c \PolEffect }{ \omegaA }
\frac{\Delta \Gamma_2}{\Delta^2 + \Gamma_2^2}. \hspace{6mm}
\label{eq:angular frequency derivative}
\end{eqnarray}

To express this in terms of linear frequencies $\nu$ we use $\delta \omegares = 2 \pi \delta \nu_\mathrm{res}$, $\omega = 2 \pi \nu$,  $\omegaA = 2 \pi \nu_\mathrm{at}$, $\Delta = 2 \pi (\nu - \nu_\mathrm{at})$, $\Gamma_2 = 2\pi \Delta \nu_\mathrm{opt}\supDtwo/2$, where $\Delta \nu_\mathrm{opt}\supDtwo$ is the FWHM of absorption in linear frequency. Making these substitutions we find the shift in resonance frequency
\begin{equation}
\delta \nu_\mathrm{} =
\frac{\pm \OD_0 c \PolEffect}{4 \pi \Lcav }
\frac{\nu}{\nu_\mathrm{at} }
\frac{(\Delta \nu_\mathrm{opt}\supDtwo/2)(\nu-\nu_\mathrm{at})}{(\nu-\nu_\mathrm{at})^2 + (\Delta \nu_\mathrm{opt}\supDtwo/2)^2}  P_z,
\label{eq:LinearFrequencyShift}
\end{equation}
where we have defined the degree of polarization $P_z=2 S_z$, equal to 1 for maximum spin polarization \cite{SeltzerThesis,Shah2009}.

\section{PDH error signal }
\label{sec:PDHSignalGeneration}
Due to phase modulation with amplitude $\beta$, the amplitude of the probe field arriving to the cavity is
\begin{eqnarray}
\cE_\mathrm{in}(t) &=& 
\sqrt{\PLaser} e^{-i \left[\omegac t + \beta \sin(\Omega t) \right]} 
\nonumber \\ & = & 
\sqrt{\PLaser} \sum_{k=-\infty}^{k=\infty} J_k(\beta) e^{-i \omega_k t },
\end{eqnarray}
where $\PLaser$ is the laser power, $J_n$ is the Bessel function of the first kind and order $n$, $\omega_k \equiv \omegac + k \Omega$, and the second equality follows from the Jacobi-Anger expansion.  

The light reflected from the cavity has amplitude
\begin{equation}
\cE_\mathrm{ref}(t) = \sqrt{\PLaser} \sum_{k=-\infty}^{\infty} \rcav(\omega_k) J_k(\beta) e^{-i \omega_k t },
\end{equation}
where $\rcav(\omega)$ is the cavity reflection amplitude of \autoref{eq:CavityRAmpl}. 

The mean power arriving to the photodetector is  
\begin{eqnarray}
\label{eq:MeanReflPower2}
\bar{P}_\mathrm{ref}(t) &=& |\cE_\mathrm{ref}(t)|^2 =
\nonumber \\ & = & 
\PLaser \sum_{k=-\infty}^{\infty} \sum_{k'=-\infty}^{\infty}  J_k(\beta) J_{k'}(\beta) \rcav^*(\omega_k) \rcav(\omega_{k'}) \nonumber \\ & & \times e^{-i (\omega_{k'} - \omega_k) t },
\end{eqnarray}
which has a dc component \begin{eqnarray}
\label{PDsignal_general}
\bar{P}_{\mathrm{ref},0}(t) &=& \PLaser \sum_{k=-\infty}^{\infty} \rcav^*(\omega_k) J_k(\beta) e^{i \omega_k t } \nonumber \\ &\times& \sum_{k'=-\infty}^{\infty} \rcav(\omega_{k'}) J_{k'}(\beta) e^{-i \omega_{k'} t } \delta_{k',k}
\nonumber \\ & = & 
\PLaser \sum_{k=-\infty}^{\infty}  |\rcav(\omega_k) J_k(\beta)|^2.
\end{eqnarray}
and a component oscillating at frequency $\Omega$

\begin{eqnarray}
\label{PDsignal_general}
\hspace{-5mm}\bar{P}_{\mathrm{ref},\Omega}(t)&=&\PLaser \sum_{k=-\infty}^{\infty} \sum_{k'=-\infty}^{\infty} \rcav^*(\omega_k) J_k(\beta) e^{i \omega_k t }  \nonumber \\ & & \times  \rcav(\omega_{k'}) J_{k'}(\beta) e^{-i \omega_{k'} t } (\delta_{k',k+1} + \delta_{k'+1,k}) \nonumber \\ & = & 
\PLaser \sum_{k=-\infty}^{\infty} J_k(\beta)J_{k+1}(\beta) \nonumber \\ & & \times 2\mathrm{Re}\left[ \rcav^*(\omega_k)  \rcav(\omega_{k+1})  e^{-i \Omega t }   \right].
\end{eqnarray}

When demodulated with phase $\varphi$, and with system gain $K$, the resulting error signal is 
\begin{equation}
\label{Eq:ErrorSignal_general}
    \epsilon =
K \sum_{k=-\infty}^{\infty} J_k(\beta)J_{k+1}(\beta)  2\mathrm{Re}\left[ \rcav^*(\omega_k)  \rcav(\omega_{k+1} )  e^{i \varphi}  \right].
\end{equation}

\bibliography{biblio}

\end{document}